\documentstyle[twoside,fleqn,espcrc2,epsfig]{article}

\newcommand{\AmS}{{\protect\the\textfont2
  A\kern-.1667em\lower.5ex\hbox{M}\kern-.125emS}}

\newcommand{\f}{\frac}
\newcommand{\be}{\begin{equation}}
\newcommand{\ee}{\end{equation}}
\newcommand{\bea}{\begin{eqnarray}}
\newcommand{\eea}{\end{eqnarray}}
\newcommand{\noi}{\noindent}
\newcommand{\pss}{\protect\scriptscriptstyle}
\newcommand{\pst}{\protect\textstyle\scriptscriptstyle}
\newcommand{\hq}{\hat{q}}

\title{The fixed point action of the Schwinger model\thanks{Work supported by 
Fondazione ``A. Della Riccia'' (Italy) and Ministerio de Educaci\'on y 
Cultura (Spain).}\thanks{Presented by F. Farchioni.}}

\author{F. Farchioni and V. Laliena\address{Institute for Theoretical Physics, 
University of Bern, \\ Sidlerstrasse 5, CH-3012 Bern, Switzerland}}

\begin{document}
\pagestyle{empty}

\begin{abstract}
We compute the fixed point action for the Schwinger model
through an expansion in the gauge field. The calculation allows a check
of the locality of the action. We test its perfection 
by computing the 1-loop mass gap at finite spatial volume. 
\end{abstract}

\maketitle

\section{INTRODUCTION}

The Wilson's Renormalization Group (RG) ensures the existence of perfect
actions, i.e. actions which reproduce the continuum
independently of the value of the lattice spacing. 
A feasible objective is the determination 
of classically perfect actions, which eliminate the cut off effects 
with restriction to the classical properties of the theory. 
These actions are related~\cite{hn0} to 
the fixed point (FP), lying on the critical surface, of a given 
RG transformation.
Although not perfect, the FP actions represent a huge step toward 
the elimination of the cut off effects~\cite{hn0,hnymn,aless}.

The final objective is the application of these ideas to QCD,
and steps in this direction have been already done~\cite{hnymn,wnf,kunszt}.
The introduction of the interaction between gauge fields 
and fermions in the context of the FP actions  
has to cope with several technical 
problems. Here we study the fermion-gauge field FP 
interactions in a case much simpler than QCD, the Schwinger model.
Since its gauge group is abelian, it is 
possible to formulate the lattice regularization with non-compact 
gauge fields, allowing the analytical solution of the pure gauge sector.
Therefore, we are able to concentrate the numerical effort in the 
fermion problem. We test the perfection of the FP action 
in perturbation theory by computing the 1-loop mass gap in a finite volume,
a circle of length $L$, using the standard action as a ``control'' action.
A non-perturbative approach to the same problem can be found in the
contribution of C.B.~Lang and T.K.~Pany to this volume~\cite{lang}.

\section{THE FP ACTION}

A general form of the action of a lattice-regularized abelian gauge theory is,
in the non-compact  formulation:
\be
S\;=\;\beta\,S_g(A)\:+\:\bar\psi\,\Delta(U)\,\psi\, ;
\ee
\noi
the fermion matrix $\Delta$ depends on the gauge field through the 
link variable $U_{\mu}=\exp(iA_{\mu})$. In the classical limit $\beta\rightarrow\infty$
the RG transformation assumes the form:
\bea
S_g^\prime(A^\prime)\;=\;\min_{\{ A \} }\,\left[S_g(A)\:+\:
\kappa_g\,{\cal K}_g (A^\prime,A)\,\right] \, ,  \label{sapo} \\
S_F^\prime (\bar\psi^\prime,\psi^\prime,U^\prime)\;=
\;\;\;\;\;\;\;\;\;\;\;\;\;\;\;\;\;\;\;\;\;\;\;\;\;\;\;\;\;\;\;\;\;\;\;\;\,\,\,
\nonumber\\
-\ln\,\int\,\left[d\bar\psi\,d\psi\right]\,\exp\,\left[\,
-\:\bar\psi\,\Delta(U(A^{\pst min}))\,\psi \right. \nonumber \\
\left.-\:\kappa_F\,{\cal K}_F(\bar\psi^\prime,\psi^\prime,
\bar\psi,\psi,U(A^{\pst min}))\,\right]\, , \label{sapof}
\eea
where the primed fields are the degrees of freedom on the coarser lattice,
defined through the gauge invariant kernels \ ${\cal K}_g$ \ and \ 
${\cal K}_F$.

In the case of the Schwinger model (and in general for asymptotically
free theories) the limit $\beta\rightarrow\infty$ is critical. The 
RG transformation of  Eqs.~(\ref{sapo}) and (\ref{sapof}) 
has a fixed point defined  by the FP 
pure-gauge action  $S^{\pss FP}_g$ and fermion matrix $\Delta^{\pss FP}$.
If ${\cal K}_F$ is quadratic in the fermion fields, the FP action remains
a bilinear form of the fermion fields: 
\be
S^{\pss FP}\;=\;\beta\,S^{\pss FP}_g(A)\:+\:\bar\psi\,\Delta^{\pss FP}(U)\,\psi\, .
\ee

We choose the simplest kernel for the gauge field:
\bea
{\cal K}_g(A^\prime,A)\;=\;\;\;\;\;\;\;\;\;\;\;\;\;\;\;\;\;\;\;\;\;\;\;\;\;\;\;\;\;\;\;\;\;\;\;\;\;\;\;\;\;\;\;\;\;\;\;\;\nonumber\\
\sum_{x_B,\mu}\!\!\left[A^\prime_\mu(x_B)\!-\!
\f{1}{2}\!\left(A_\mu(2x_B)\!+\!
A_\mu(2x_B\!+\!\hat\mu)\right)\right]^2\! .
\label{gaugekernel}
\eea
\noi
The variables $x_B$ label the sites on the coarse lattice in units of 
its doubled lattice spacing: the site $x_B$ corresponds to the site
$2x_B$ in the units of the original lattice.
Taking $\kappa_g\rightarrow\infty$ we recover the original standard 
non-compact action as FP action, which is ultralocal, involving only 
nearest-neighbors interactions.

The fermion kernel has the form:
\bea
{\cal K}_F(\bar\psi^\prime,\psi^\prime,\bar\psi,\psi,U)\;=
\;\;\;\;\;\;\;\;\;\;\;\;\;\;\;\;\;\;\;\;\;\;\;\;\;\;\;\;\;\;\;\;\;\nonumber\\
\sum_{x_B}\!
\left[\bar\psi^\prime(x_B)\!-\!\bar\Gamma(x_B,U)\right]\!
\left[\psi^\prime(x_B)\!-\!\Gamma(x_B,U)\right] .
\eea
The function $\Gamma(x_B,U)$ defines a gauge-invariant average of the
fine fields. Only local, first and second neighbors fine fields contribute 
to a given coarse field, each with a weight inversely proportional 
to the total number of coarse fields to which it contributes. 
Gauge invariance is achieved by parallel-transporting the fine fields 
to the site $x_B$ through the shortest path. 

An additional factor-two renormalization of the gauge field $A_\mu$
is required when introducing the interaction, since in $d=2$ the electric 
charge $e$ has the dimension of a mass.

We solve the recursive equation (\ref{sapof})
in an expansion
in the gauge field $A_\mu$. As a result, we find the FP first and second
order vertex $R^{\pss (1)}_\mu$ and $R^{\pss (2)}_{\mu\nu}$, 
defined by (summation over repeated indices is understood):
\bea
\Delta^{\pss FP}(x,x',U(A)) =\;\;\;\;\;\;\;\;\;\;\;\;\;\;\;\;\;\;\;\;\;\;\;\;\;\;\;\;\;\;\;\;\;\;\;\;\;\;\;\;\;\; \nonumber \\
D^{-1}(x-y)\,+\,i\,R^{\pss (1)}_\mu(x-r,y-r)\,A_\mu(r) + \;\;\;\;\nonumber \\
R^{\pss (2)}_{\mu\nu}(x-r,y-r',r-r')A_\mu(r)\,A_\nu(r')+\ldots
\label{sexpan}
\eea
$D(x-y)$ stands for the perfect fermion propagator for the free-field case.

\begin{figure}[htb]
\mbox{\psfig{file=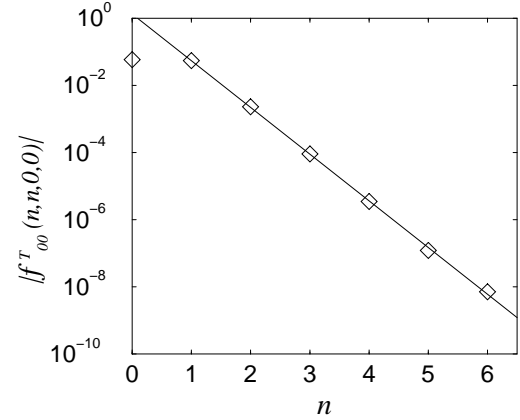,width=6 truecm, angle=-90}}
\caption{Exponential decay of $|f^{\pss T}_{00}(n,n,0,0)|$, defined by
$f^{\pss T}_{00}=\f{1}{2}\,{\rm Tr}(\gamma_0R^{\pss (1)}_0)$;
the continuum line is the best fit $\sim\exp(-n/0.44)$.}
\label{fig:dec}
\end{figure}

The FP couplings decay exponentially with the distance (see
Fig.~\ref{fig:dec}), the most important being those 
coupling fields up to second neighbors. 
From this observation we conclude that a good parametrization
of the full FP action should be possible by using couplings confined 
in a $2\times2$ plaquette.

\section{THE MASS GAP}

The mass gap of the lattice theory is given by the solution
of the equation:
\be
\left|\hq_0\right|^2\:-\:\Pi_{11}(q_0,q_1=0)\;=\;0 \, , \label{masseq}
\ee
where $\hq_\mu = \exp(iq_\mu) - 1$ and  $\Pi_{\mu\nu}(q)$ is the 
lattice vacuum polarization tensor.
The perturbative expansion of $\Pi_{\mu\nu}$
defines an analogous expansion for the mass gap; considering 
the model defined on a cylinder of (spatial) circumference $L$:
\be
m(g,a/L)=g\,m^{\pss (1)}(a/L)+
g^3\,m^{\pss (2)}(a/L)+\ldots\;\label{pmass}
\ee
The scaling limit $a\rightarrow 0$, with $g=ea$,
gives the continuum mass $m_{\pss ph}^{\pss (c)}=\frac{e}{\sqrt{\pi}}$:
\bea
\frac{1}{e}\,m_{\pss ph}(ea,a/L)\,=\,m^{\pss (1)}(a/L)\,+
\;\;\;\;\;\;\;\;\;\;\;\;\;\,\nonumber \\
e^2a^2\,m^{\pss (2)}(a/L)\,+\,\ldots 
\;\rightarrow\;
\frac{m_{\pss ph}^{\pss (c)}}{e}\;=\;\frac{1}{\sqrt\pi}\,\; .
\label{cont}
\eea

\begin{figure}[htb]
\mbox{\psfig{file=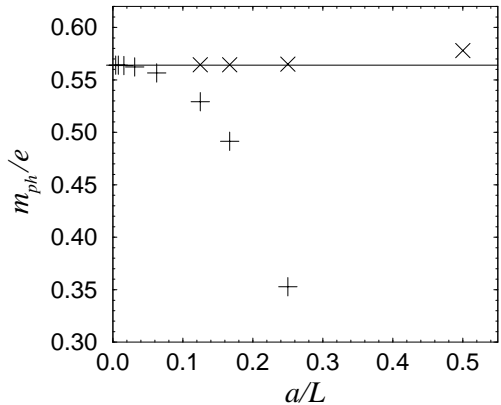,width=6 truecm, angle=-90}}
\caption{The ratio $m_{\pss ph}/e$ as a function of $a/L$ for the
Wilson action (plus) and the FP action (crosses); the solid line is 
the continuum value $1/\sqrt{\pi}$.}
\label{fig:mass}
\end{figure}

We calculated the lowest order term $m^{\pss (1)}(a/L)$,
where only the first and second order vertex enter;
higher order terms are pure cut off effects.
In Fig.~\ref{fig:mass} we report the results
for the Wilson (non-compact) action and the FP action.
With the Wilson action we see clear power-like cut off effects, according to
the law $m_{\pss ph}/e  = 0.5641900 + 1.9\cdot(a/L)^2 + O((a/L)^4)$; for
$L/a=2$ the lattice theory contains no particle at all. In the case
of the FP action, for $L/a>2$, only tiny ($O(10^{-4})$) deviations 
from the continuum value are observed (we attribute these deviations 
to the numerical approximation in the determination of the FP vertices).
The deviation for $L/a=2$ is related to an additional effect~\cite{fphn}
exponentially decaying when $L/a$ increases and related to the finite 
extension of the FP action.

\section{1-LOOP PERFECTION}

The recent debate~\cite{hn1l} about the 1-loop perfection of the 
FP action, supported by a formal RG argument, but  
disproved~\cite{hn1l} by an explicit calculation in the case 
of the non-linear O(3)-$\sigma$ model, urged us to consider whether
our result can represent or not a check of this property. 
The property of 1-loop perfection of the FP action is stated in terms
of its behavior under RG transformations at finite $\beta$. After one RG step: 
\bea
S^{\pss FP}\;\rightarrow\;\frac{\beta}{4}\,(\,S_g^{\pss FP}(A)\:+\:\delta
S_g(A,\beta)\,)\:+\nonumber\\
\bar\psi\,\Delta^{\pss FP}(U)\,\psi
\:+\:\delta S_{\pss F}(\bar\psi,\psi,U,\beta)\;;
\label{ass}
\eea
1-loop perfection means absence of the lowest-order ($O(1/\beta)$) 
corrections to the self-reproducing behavior of the FP action
under the RG transformation.
After the inspection of all possible
effective vertices contributing to these lowest-order corrections,
we realize that none of them (even if present) would affect the
1-loop mass gap. As a consequence, we conclude that our result represents 
indeed a check of the {\em classical} perfection of the FP action
of the Schwinger model. A similar situation is found in the framework 
of the O(3)-$\sigma$ model. 

We thank P.~Hasenfratz and F.~Niedermayer for having
introduced us in the subject and for useful suggestions.

\end{document}